\documentclass[useAMS,usenatbib]{mn2e}

\usepackage{amssymb}
\usepackage{graphicx}

\usepackage{color,soul}

\def\aj{AJ}%          % Astronomical Journal
\def\araa{ARA\&A}%          % Annual Review of Astron and Astrophys
\def\apj{ApJ}%          % Astrophysical Journal
\def\apjl{ApJ}%          % Astrophysical Journal, Letters
\def\apjs{ApJS}%          % Astrophysical Journal, Supplement
\def\aap{A\&A}%          % Astronomy and Astrophysics
\def\aaps{A\&AS}%          % Astronomy and Astrophysics, Supplement
\def\mnras{MNRAS}%          % Monthly Notices of the RASmega
\def\na{New A}%          % New Astronomy
\def\pasa{PASA}%          % Publications of the Astron. Soc. of Australia
\def\nat{Nature}%          % Nature
\title[UCD formation]{Ultra compact dwarf galaxy formation by tidal stripping of nucleated dwarf galaxies}
\author[J. Pfeffer and H. Baumgardt]{J. Pfeffer$^{1}$\thanks{E-mail: j.pfeffer@uq.edu.au} and H. Baumgardt$^{1}$\\
$^{1}$School of Mathematics and Physics, The University of Queensland, Brisbane, QLD 4072, Australia}

\begin{document}

\date{}

\pagerange{\pageref{firstpage}--\pageref{lastpage}} \pubyear{2012}

\maketitle

\label{firstpage}

%%%%%%%%%%%%%%%%%%%%%%%%%%%%%%%%%%%%%%%%%%
%                ABSTRACT                %
%%%%%%%%%%%%%%%%%%%%%%%%%%%%%%%%%%%%%%%%%%
\begin{abstract}
Ultra Compact Dwarf Galaxies (UCDs) and dwarf galaxy nuclei have many common properties, such as 
internal velocity dispersions and colour-magnitude trends, suggesting tidally stripped dwarf 
galaxies as a possible UCD origin. 
However, UCDs typically have sizes more than twice as large as nuclei at the same luminosity. 
We use a GPU-enabled version of the particle-mesh code \textsc{superbox} to study the possibility of 
turning nucleated dwarf galaxies into UCDs by tidally stripping them in a Virgo-like galaxy cluster. 
We find that motion in spherical potentials, where close passages happen many times, leads to the 
formation of compact ($r_h \lesssim 20$ pc) star clusters/UCDs. 
In contrast, orbital motion where close passages happen only once or twice leads to the formation of 
extended objects which are large enough to account for the full range of observed UCD sizes. 
For such motion, we find that dwarf galaxies need close pericentre passages with distances less than 
10 kpc to undergo strong enough stripping so that UCD formation is possible. 
As tidal stripping produces objects with similar properties to UCDs, and our estimates suggest dwarf 
galaxies have been destroyed in sufficient numbers to explain the observed number of UCDs in M87, we 
consider tidal stripping to be a likely origin of UCDs. 
However, comparison with cosmological simulations is needed to determine if the number and spatial 
distribution of UCDs formed by tidal stripping matches the observations of UCDs in galaxy clusters.
\end{abstract}


%%%%%%%%%%%%%%%%%%%%%%%%%%%%%%%%%%%%%%%%%%
%              INTRODUCTION              %
%%%%%%%%%%%%%%%%%%%%%%%%%%%%%%%%%%%%%%%%%%
\section{INTRODUCTION}

Ultra compact dwarf galaxies (UCDs) are a class of stellar systems that was discovered in the Fornax 
galaxy cluster more than a decade ago \citep{Hilker:1999a, Drinkwater:2000}. 
Since then, UCDs have been discovered in other galaxy clusters \citep{Mieske:2004, Hasegan:2005, 
Jones:2006, Mieske:2007, Misgeld:2011, Madrid:2011, Penny:2012}, galaxy groups 
\citep{Evstigneeva:2007b, DaRocha:2011}, as well as isolated spiral galaxies \citep{Hau:2009}. 
UCDs are typically defined to have half-light radii $7 \lesssim r_h / \rmn{pc} \lesssim 100$ and 
masses $M \gtrsim 2 \times 10^6$ M$_\odot$ \citep{Mieske:2008}, making them an intermediate object 
between globular clusters (GCs) and dwarf galaxies. 

The formation mechanism of UCDs is currently unknown, although a number of scenarios have been 
proposed. 
The simplest explanation is that they are the high-mass end of the GC mass function observed around 
galaxies with rich GC systems \citep*{Mieske:2002, Mieske:2012}. 
Since UCDs have larger sizes than typical GCs, they may form from the merger of many GCs 
\citep{Kroupa:1998, Fellhauer:2002, Bruens:2011, Bruens:2012}. 
Alternatively, they could be dwarf galaxies stripped by tidal interactions such that only their 
central nuclei remain, referred to as the tidal stripping or ``threshing'' scenario 
\citep*{Bassino:1994, Bekki:2001, Bekki:2003, Drinkwater:2003}. 
There is also evidence suggesting UCDs are formed by a combination of mechanisms rather than a 
single one \citep{Mieske:2006, Brodie:2011, Chilingarian:2011, Norris:2011}.

Tidally stripped dwarf galaxies, in particular nucleated dwarf elliptical galaxies (dE,Ns), have 
been proposed as UCD progenitors for a number of reasons: UCDs and dwarf elliptical nuclei have 
similar internal velocity dispersions \citep{Drinkwater:2003} and colour-magnitude trends 
\citep{Cote:2006, Evstigneeva:2008, Brodie:2011}. 
UCDs lie above the metallicity-luminosity trend for early-type galaxies and have a similar 
metallicity to dwarf galaxies \citep{Chilingarian:2011, Francis:2012}. 
Such a situation would be expected if UCDs are tidally stripped dwarf galaxies since the luminosity 
would decrease while the metallicity remains high. 
In addition, some UCDs are surrounded by stellar haloes which might be the remnants of the stripped 
dwarf galaxies \citep{Drinkwater:2003, Hasegan:2005, ChilingarianMamon:2008, Evstigneeva:2008, 
Chiboucas:2011}. 
Finally, irregular objects with asymmetric extensions have been found which may be dwarf galaxy 
nuclei undergoing tidal stripping \citep{Richtler:2005, Brodie:2011}.

Simulations of UCD formation in the tidal stripping scenario were first performed by 
\citet{Bekki:2001, Bekki:2003}, who showed that nucleated dwarf galaxies orbiting in a galaxy 
cluster on highly eccentric orbits are almost completely tidally stripped, with only the nucleus 
surviving. 
In this case the size and luminosity of the stripped galaxy are such that it would be classified as 
a UCD. 
More recently, \citet{Goerdt:2008} sought to understand UCD formation by tidal stripping within the 
context of the cold dark matter (CDM) model, and found the cosmological prediction matches the 
observed spatial distribution of UCDs.

Despite this, a number of potential problems with the tidal stripping scenario still remain 
unanswered. 
\citet{Brodie:2011} have recently found a population of extended, low mass UCDs ($r_h \sim 40$ pc, 
$M_V \sim -9$) in the Virgo galaxy cluster which do not follow the size-magnitude relation observed 
for the most massive UCDs \citep{Mieske:2006, Evstigneeva:2008}. 
Such objects may form from merged star cluster complexes \citep{Bruens:2011, Bruens:2012}, however 
it is unclear whether they can be formed in either the giant GC or tidal stripping scenario.

UCDs are typically $\sim 2$ times larger than dE nuclei at the same luminosity 
\citep{Evstigneeva:2008}. 
This suggests that if nuclei are the UCD progenitors the nuclei must undergo expansion, or some 
other process, during tidal stripping. 
The situation is even more pronounced for nuclei and UCDs with $M_V > -11$: at this luminosity the 
typical half-light radius of nuclei is $\sim 4$ pc, while UCDs extend up to $\sim 40$ pc 
\citep{Brodie:2011}.

All tidal stripping simulations to date have been performed in static, spherical galaxy potentials 
with a constant pericentre and apocentre \citep{Bekki:2001, Bekki:2003, Goerdt:2008}, thus it is 
uncertain what effect an evolving or triaxial potential has on the number and properties of the UCDs 
that form in this scenario. 
In addition, these studies did not investigate if tidal stripping of dwarf galaxies affects the 
nucleus size, thus the origin of the size difference between UCDs and dE nuclei is still uncertain.

In this paper we perform high resolution $N$-body simulations of a nucleated dwarf galaxy being 
stripped via tidal interactions to test whether the extended, low mass UCDs can be formed in the 
tidal stripping scenario. 
We simulate orbits with fixed pericentres and apocentres, as well as orbits that mimic `box orbits' 
in an evolving or triaxial potential, to study the effect of the dwarf galaxy's orbit on the final 
size of the stripped dwarf galaxy.

\begin{table*}
\centering
\begin{minipage}{168mm}
\centering
\caption{Parameters of the initial dwarf galaxy (columns 2-5) and the resulting UCD (columns 6-8) for all simulations. 
dE,N model 1 has a nucleus half-light radius $r_h = 4$ pc and magnitude $M_V = -10$ (nucleus mass $M_\rmn{nuc} = 2.56 \times 10^6$ M$_\odot$). 
dE,N model 2 has a nucleus half-light radius $r_h = 10$ pc and magnitude $M_V = -10$ (nucleus mass $M_\rmn{nuc} = 2.56 \times 10^6$ M$_\odot$). 
dE,N model 3 has a nucleus half-light radius $r_h = 10$ pc and magnitude $M_V = -12$ (nucleus mass $M_\rmn{nuc} = 1.62 \times 10^7$ M$_\odot$). 
Masses are converted to magnitudes assuming a mass-to-light ratio $\Upsilon = 3$ $(\rmn{M}/\rmn{L}_V)_\odot$.}
\label{results_table}  
\begin{tabular} {@{}ccccccccccc@{}}
  \hline
  Simulation & Orbit & dE,N & Apocentre & Pericentre & Close & Final & Final & Formation & Simulation \\
  no. & type & Model & (kpc) & (kpc) & pericentre & $r_h$ & $M_V$ & time & time \\
  & & & & & passages & (pc) & (mag) & (Gyr) & (Gyr) \\  
  \hline
  1  & Elliptic & 1 & $\;\:$50 & $\;\:$2 & $\;\:$5 & $\quad$3.9 & $\;\:$-9.77 & 0.9 & 2.2 \\
  2  & Elliptic & 1 & $\;\:$50 & $\;\:$5 & $\;\:$7 & $\quad$5.5 &      -10.10 & 1.3 & 2.2 \\
  3  & Elliptic & 1 & $\;\:$50 &      10 &      12 & $\quad$6.6 &      -10.12 & 2.3 & 4.2 \\
  4  & Elliptic & 1 & $\;\:$50 &      20 &      27 & $\quad$8.6 &      -10.14 & 5.7 & 7.5 \\
  5  & Elliptic & 1 &      100 & $\;\:$2 & $\;\:$5 & $\quad$4.2 & $\;\:$-9.71 & 1.6 & 3.9 \\
  6  & Elliptic & 1 &      100 & $\;\:$5 & $\;\:$9 & $\quad$6.1 &      -10.07 & 2.9 & 3.9 \\
  7  & Elliptic & 1 &      100 &      10 &      15 & $\quad$7.6 &      -10.14 & 5.0 & 5.9 \\
  8  & Elliptic & 2 & $\;\:$50 & $\;\:$2 & $\;\:$6 & $\quad$5.6 & $\;\:$-9.02 & 1.1 & 2.2 \\
  9  & Elliptic & 2 & $\;\:$50 & $\;\:$5 & $\;\:$7 & $\quad$9.5 & $\;\:$-9.86 & 1.3 & 2.2 \\
  10 & Elliptic & 2 & $\;\:$50 &      10 &      11 & $\;\:$11.7 &      -10.02 & 2.1 & 4.2 \\
  11 & Elliptic & 2 & $\;\:$50 &      20 &      25 & $\;\:$18.7 &      -10.32 & 5.2 & 7.5 \\
  12 & Elliptic & 2 &      100 & $\;\:$2 & $\;\:$6 & $\quad$5.9 & $\;\:$-9.07 & 1.9 & 3.9 \\
  13 & Elliptic & 2 &      100 & $\;\:$5 & $\;\:$9 & $\;\:$10.3 & $\;\:$-9.90 & 2.9 & 3.9 \\
  14 & Elliptic & 2 &      100 &      10 &      17 & $\;\:$15.2 &      -10.19 & 5.6 & 5.9 \\
  15 & Elliptic & 3 & $\;\:$50 & $\;\:$2 & $\;\:$5 & $\quad$7.0 &      -11.43 & 0.9 & 2.2 \\
  16 & Elliptic & 3 & $\;\:$50 & $\;\:$5 & $\;\:$6 & $\;\:$12.0 &      -12.01 & 1.1 & 2.2 \\
  17 & Elliptic & 3 & $\;\:$50 &      10 & $\;\:$9 & $\;\:$15.5 &      -12.13 & 1.7 & 4.2 \\
  18 & Elliptic & 3 & $\;\:$50 &      20 &      15 & $\;\:$24.3 &      -12.30 & 3.1 & 7.5 \\
  19 & Elliptic & 3 &      100 & $\;\:$2 & $\;\:$5 & $\quad$7.7 &      -11.31 & 1.6 & 3.9 \\
  20 & Elliptic & 3 &      100 & $\;\:$5 & $\;\:$6 & $\;\:$13.9 &      -12.04 & 1.9 & 3.9 \\
  21 & Elliptic & 3 &      100 &      10 &      11 & $\;\:$19.3 &      -12.16 & 3.6 & 5.9 \\
  22 & Box      & 1 & $\;\:$50 & $\;\:$2 & $\;\:$1 & $\;\:$28.3 &      -10.84 & 2.0 & 2.2 \\
  23 & Box      & 1 & $\;\:$50 & $\;\:$2 & $\;\:$2 & $\quad$7.7 &      -10.30 & 1.3 & 2.2 \\
  24 & Box      & 1 & $\;\:$50 & $\;\:$5 & $\;\:$2 & $\;\:$49.8 &      -11.10 & 1.1 & 2.2 \\
  25 & Box      & 1 & $\;\:$50 & $\;\:$5 & $\;\:$3 & $\;\:$14.6 &      -10.61 & 1.1 & 2.2 \\
  26 & Box      & 1 &      100 & $\;\:$2 & $\;\:$1 & $\;\:$94.1 &      -11.36 & 1.9 & 3.9 \\
  27 & Box      & 1 &      100 & $\;\:$2 & $\;\:$2 & $\quad$9.7 &      -10.33 & 2.2 & 3.9 \\
  28 & Box      & 1 &      100 & $\;\:$5 & $\;\:$2 &      179.0 &      -12.22 & 1.6 & 3.9 \\
  29 & Box      & 1 &      100 & $\;\:$5 & $\;\:$3 & $\;\:$58.1 &      -11.11 & 2.6 & 3.9 \\
  30 & Box      & 2 & $\;\:$50 & $\;\:$2 & $\;\:$1 & $\;\:$46.3 &      -10.86 & 1.2 & 2.2 \\
  31 & Box      & 2 & $\;\:$50 & $\;\:$2 & $\;\:$2 & $\;\:$13.4 &      -10.09 & 1.4 & 2.2 \\
  32 & Box      & 2 & $\;\:$50 & $\;\:$5 & $\;\:$2 & $\;\:$59.6 &      -11.10 & 0.9 & 2.2 \\
  33 & Box      & 2 & $\;\:$50 & $\;\:$5 & $\;\:$3 & $\;\:$25.1 &      -10.58 & 1.1 & 2.2 \\
  34 & Box      & 2 &      100 & $\;\:$2 & $\;\:$1 & $\;\:$98.2 &      -11.37 & 1.9 & 3.9 \\
  35 & Box      & 2 &      100 & $\;\:$2 & $\;\:$2 & $\;\:$17.1 &      -10.21 & 2.2 & 3.9 \\
  36 & Box      & 2 &      100 & $\;\:$5 & $\;\:$2 &      168.0 &      -12.13 & 1.7 & 3.9 \\
  37 & Box      & 2 &      100 & $\;\:$5 & $\;\:$3 & $\;\:$65.8 &      -11.12 & 2.0 & 3.9 \\
  38 & Box      & 3 & $\;\:$50 & $\;\:$2 & $\;\:$1 & $\;\:$28.7 &      -12.50 & 1.2 & 2.2 \\
  39 & Box      & 3 & $\;\:$50 & $\;\:$2 & $\;\:$2 & $\;\:$15.0 &      -12.15 & 0.5 & 2.2 \\
  40 & Box      & 3 & $\;\:$50 & $\;\:$5 & $\;\:$2 & $\;\:$32.1 &      -12.59 & 1.3 & 2.2 \\
  41 & Box      & 3 & $\;\:$50 & $\;\:$5 & $\;\:$3 & $\;\:$20.5 &      -12.38 & 0.7 & 2.2 \\
  42 & Box      & 3 &      100 & $\;\:$2 & $\;\:$1 & $\;\:$65.8 &      -12.80 & 1.9 & 3.9 \\
  43 & Box      & 3 &      100 & $\;\:$2 & $\;\:$2 & $\;\:$18.7 &      -12.21 & 1.0 & 3.9 \\
  44 & Box      & 3 &      100 & $\;\:$5 & $\;\:$2 &      118.0 &      -13.08 & 2.2 & 3.9 \\
  45 & Box      & 3 &      100 & $\;\:$5 & $\;\:$3 & $\;\:$37.7 &      -12.60 & 1.6 & 3.9 \\
    \hline
\end{tabular}
\end{minipage}
\end{table*}

%%%%%%%%%%%%%%%%%%%%%%%%%%%%%%%%%%%%%%%%%%
%                 MODELS                 %
%%%%%%%%%%%%%%%%%%%%%%%%%%%%%%%%%%%%%%%%%%
\section{THE SIMULATIONS}

\subsection{Simulation code}

All simulations were performed with a graphics processing unit (GPU) enabled version of the 
particle-mesh code \textsc{superbox} \citep{Fellhauer:2000} on NVIDIA GPUs. 
\textsc{superbox} uses a leap-frog scheme to integrate the motion of particles and has 
high-resolution subgrids which stay focused on the simulated objects. 
In \textsc{superbox} the density grids are derived using a nearest-grid-point (NGP) scheme. 
Potentials are calculated from the density grids using a Fast Fourier Transformation, which are 
performed in parallel across the subgrids using multiple GPUs. 
The forces are calculated using a NGP scheme based on the second derivatives of the potential. 
We have increased the subgrid number from 2 to 4 in order to accurately resolve the innermost 
regions of our dE,N models and reduce edge effects for particles crossing between subgrids. 
For the simulations we use $64^3$ grid points for all subgrids with an innermost subgrid size of 
0.013 kpc for model 1 and 0.025 kpc for models 2 and 3, and subsequent grid sizes of 0.05, 0.5, 4 
and 40 kpc for all models.
Time steps are chosen such that no particle moves more than a grid cell length of the smallest 
subgrid per time step. We use time steps of 0.004 Myr, 0.008 Myr and 0.006 Myr for dE,N models 1, 2 
and 3, respectively.

%%%%%%%%%%%%%%%%%%%%%%%%%%%%%%%%%%%%%%%%%%
\subsection{Cluster mass profile}

In order to have a realistic mass profile for the galaxy cluster, we model M87, the central galaxy 
of the Virgo cluster, based on the observations of \citet{Kormendy:2009} and \citet*{Murphy:2011} 
using S{\'e}rsic and logarithmic profiles for the stellar and dark matter components, respectively. 
The \citet{Sersic:1963, Sersic:1968} surface brightness profile is given by
\begin{equation} \label{sersic_sb}
I(R) = I_0 e^{-b(R/R_e)^{1/n}}
\end{equation}
where $I_0$, $R_e$ and $n$ are the central intensity, the effective half-light radius and the 
\textit{S{\'e}rsic index} describing the curvature of the profile, respectively. 
The constant $b$ is chosen such that $R_e$ contains half the projected light and we use the relation 
between $b$ and $n$ found by \citet{Prugniel:1997}. 
For the simulations the S{\'e}rsic surface brightness profile is converted into a potential using 
the method of \citet{Terzic:2005}. 
We adopt for the stellar component a S{\'e}rsic index $n = 11.84$, an effective radius $R_e = 16.22$ 
kpc and a central intensity $I_0 = 2.732 \times 10^{17}$ L$_\odot$ kpc$^{-2}$, consistent with 
\citet{Kormendy:2009}, and a mass-to-light ratio $\Upsilon = 9.1$ $(\rmn{M}/\rmn{L})_\odot$ 
consistent with \citet{Murphy:2011}. 

The logarithmic potential of the dark matter halo is given by
\begin{equation} \label{log_dens}
\phi(r) = \frac{v_c^2}{2} \ln(r_c^2 + r^2)
\end{equation}
where $r_c$ is the core radius and $v_c$ is the asymptotic circular velocity. 
We a adopt a scale radius $r_c = 36$ kpc and a circular velocity of $v_c = 800$ km s$^{-1}$ 
consistent with \citet{Murphy:2011}.

For the cluster potential we choose a static, spherical potential similar to what has been 
chosen by other authors \citep{Bekki:2001, Bekki:2003, Goerdt:2008}.
In order to add a time varying or triaxial component to the cluster potential it would be 
necessary to add more parameters to the simulations.
In addition, in order to take into account substructure in the cluster which could perturb the 
orbits of the dwarf galaxy, it would be necessary to simulate a full galaxy cluster.
Therefore, to reduce unnecessary complexity in the simulations, we mimic these possible situations 
by changing the orbit of the dE,N during the simulations. 
We discuss the dwarf galaxy orbits in more detail in Section \ref{sec:orbits}.

%%%%%%%%%%%%%%%%%%%%%%%%%%%%%%%%%%%%%%%%%%
\subsection{Nucleated dwarf ellipticals}

We construct three models for dE,Ns according to values observed for dEs in the Virgo cluster. 
The model number for each simulation is shown in column 3 of Table \ref{results_table}. 
The nucleus is modelled using a \citet{King:1962} profile with a concentration parameter $c = 1.5$, 
with the value of $c$ chosen since it is a typical value for GCs \citep*{Trager:1995} which are 
similar in size to dE nuclei. 
An absolute $V$-band magnitude $M_V = -10$ is used for the nucleus of dE,N models 1 and 2, and 
$M_V = -12$ is used for model 3. 
For dE,N model 1 we use a nucleus half-light radius $r_h = 4$ pc (the average radius at this 
luminosity, see Fig. \ref{rh-MV}), while for models 2 and 3 we use $r_h = 10$ pc. 
The magnitude $M_V = -10$ is chosen to be comparable to the low luminosity UCDs observed by 
\citet{Brodie:2011}, even though the least luminous UCDs observed by \citeauthor{Brodie:2011} 
have $M_V=-9$. 
We chose $M_V = -12$ to be comparable to a typical massive UCD. 
Note that the most massive UCDs have $M_V=-13$.

The main stellar component of the dwarf galaxy, referred to hereafter as the `envelope,' is modelled 
using a S{\'e}rsic profile. 
For dE,N models 1 and 2 we use a S{\'e}rsic index $n = 1.5$ and an effective radius $R_e = 850$ pc, 
which are the average values observed by \citet*{Grant:2005} and \citet*{Geha:2003} for dEs in the 
Virgo cluster, respectively, while for model 3 we use a S{\'e}rsic index $n = 1.5$ and an effective 
radius $R_e = 2000$ pc. 
The central intensity is chosen such that the nucleus-to-envelope luminosity ratio is $0.3\%$, the 
mean value for dEs in Virgo \citep{Cote:2006}, giving an absolute $V$-band magnitude $M_V = -16.3$ 
for dE,N models 1 and 2, and $M_V = -18.3$ for dE,N model 3.

A mass-to-light ratio $\Upsilon = 3$ $(\rmn{M}/\rmn{L}_V)_\odot$ is set for both the nucleus and 
envelope, which is consistent with the average values observed by \citet{Chilingarian:2009}. 
In order to reduce the number of particles needed, we do not include a dark matter halo for our dE,N 
models. 
We expect that including a cored dark matter profile in the models would not significantly affect 
our results for UCD sizes and masses since we are only interested in the centre of the model where 
the fraction of dark matter to stellar matter is lowest. 
This may not be the case if a cuspy dark matter profile was used since dwarf galaxies with cuspy 
dark matter profiles are more resilient to tidal stripping \citep{Bekki:2003, Penarrubia:2010}. 
However, observations suggest dE galaxies do not have a significant amount of dark matter within one 
effective radius \citep*{Geha:2002}, more consistent with a cored profile, justifying our neglect of 
a dark matter halo.

Although the simulated UCD sizes and masses would be unaffected by a dark matter halo, it is likely 
the inclusion of one would change the UCD formation time due to the dark matter shielding the 
stellar component from disruption (e.g. in Fig. \ref{sb_profile} the envelope shields the nucleus 
during tidal stripping).
Since a dark matter halo would be much more extended than the stellar component, we expect one to 
two close passages would be necessary to disrupt the halo and therefore add $\sim 0.5$ Gyr to the 
formation time.
However, as a dwarf galaxy sized halo disrupts at a radius of a few hundred kpc in a galaxy cluster 
\citep{Goerdt:2008}, it is possible that the halo is stripped before the dwarf galaxy reaches the 
centre of the cluster.

We create an $N$-body representation for the dwarf galaxies using the following method adapted from 
\citet{Hilker:2007}:
\begin{enumerate}
\item Deprojection of the 2-dimensional surface density profile by means of the Abel integral 
equation \citep[see equation 1B-57b of][]{Binney:1987} into a 3-dimensional density profile.
\item Calculation of the cumulative mass function $M(< r)$ and the potential energy $\phi(r)$ from 
the 3-dimensional density profile. From these the energy distribution function $f(E)$ is then 
calculated with the help of equation 4-140a from \citet{Binney:1987}, assuming isotropic orbits for 
the stars.
\item Creation of an $N$-body representation of the dE using the deprojected density profile and the 
distribution function. 
\end{enumerate}
The modelling is based on the assumptions of spherical symmetry and an underlying isotropic velocity 
distribution. 
For all dE,N models $10^7$ particles with equal masses were distributed, corresponding to particle 
masses of 85.7 M$_\odot$ with $2.56 \times 10^6$ M$_\odot$ contained in the nucleus for dE,N models 
1 and 2, and particle masses of 541 M$_\odot$ with $1.62 \times 10^7$ M$_\odot$ contained in the 
nucleus for model 3.

To test the stability of the models they were evolved in isolation for 2 Gyr. 
For model 2 the Lagrange radii are conserved to better than 4 per cent. 
For model 1 the half-light radius of the nucleus increases by 13 per cent due to numerical 
relaxation, however outside the central 30 pc the maximum radial change is less than 4 per cent. 
For model 3 the half-light radius of the nucleus increases by 12 per cent due to numerical 
relaxation while outside the central 45 pc the maximum radial change is less than 4 per cent.

%%%%%%%%%%%%%%%%%%%%%%%%%%%%%%%%%%%%%%%%%%
\subsection{Dwarf galaxy orbits} 
\label{sec:orbits}

We test for nucleus expansion in the tidal stripping scenario with two types of orbits. 
First we assume that dE galaxies orbit the central galaxy on elliptic orbits with fixed apocentre 
and pericentre distances. 
For all dE,N models we test orbits with apocentres of 50 and 100 kpc and pericentres of 2, 5, 10 and 
20 kpc. 
Such highly eccentric orbits are predicted by $\Lambda$CDM simulations for sub-haloes orbiting 
larger haloes \citep{Ghigna:1998}. 

Since elliptical galaxies are expected to reside in triaxial potentials and form through galaxy 
mergers, galaxy cluster potentials will be neither static nor spherical. 
The orbits of dwarf galaxies in such clusters may be chaotic with strongly varying pericentres 
distances during successive passages, either due to the triaxial potential or encounters with other 
galaxies during the orbit. 
Thus in addition to the elliptic orbit, we consider the case where only a few encounters between the 
dE and central galaxy will happen with very small pericentre distances, and at all other times the 
dE is far from the central galaxy so that tidal effects are not important. 
We mimic such a scenario by simulating 1-3 pericentre passages and then placing the object on a 
circular orbit at apocentre to allow enough time for unbound particles to escape. 
This is referred to hereafter as the `box' orbit. 
For all dE,N models we test box orbits with apocentres of 50 and 100 kpc and pericentres of 2 and 5 
kpc. 
Note we do not require that the orbits are circular (which is highly unlikely in $\Lambda$CDM), but 
only that the pericentre increases to $\gtrsim 10$-20 kpc. 

These two orbits represent the most extreme cases in the tidal stripping scenario, i.e. galaxies may 
have either many pericentre passages at small radii, or they have many passages at large radii and 
very few at small radii. 
It is probable that real UCDs will be on orbits somewhere in between these extremes.

\begin{figure*}
\centering
\begin{minipage}{168mm}
\centering
  \includegraphics[scale=0.5]{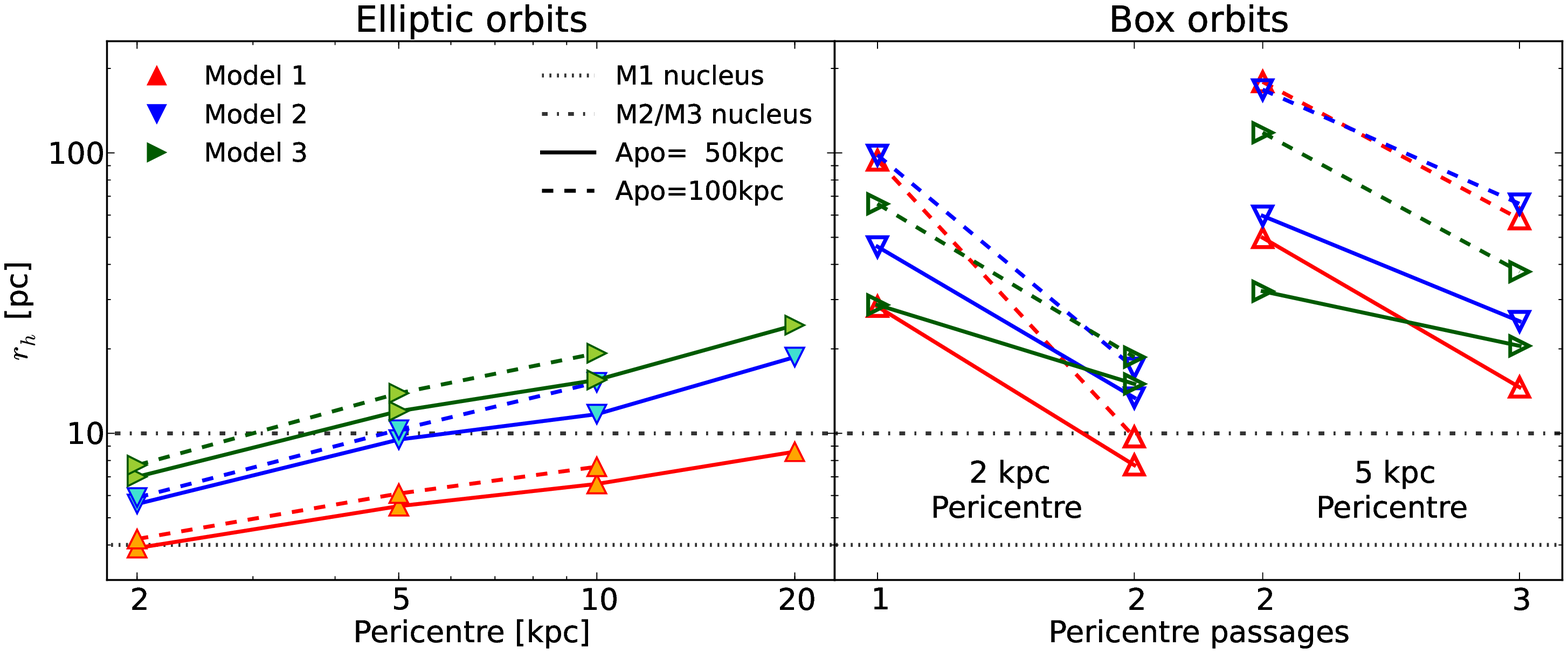}
  \caption{Comparison of the final half-light radii for all simulated UCDs in dependence of orbital 
  type, dE,N model, and apocentre distance. 
  The left panel shows the effect of pericentre distance for models on elliptic orbits, while the 
  right panel shows the effect of pericentre distance and number of close pericentre passages for 
  models on box orbits. 
  Symbols are as in the legend, where triangle-up (red), triangle-down (blue) and triangle-right 
  (green) denote dE,N model 1 (nucleus $r_h = 4$ pc, $M_V = -10$ mag), model 2 (nucleus $r_h = 10$ 
  pc, $M_V = -10$ mag) and model 3 (nucleus $r_h = 10$ pc, $M_V = -12$ mag), respectively, and 
  orbits with a 50 (100) kpc apocentre are represented by a solid (dashed) line. 
  The dotted and dash-dot lines show the initial half-light radius of the nucleus for dE,N models 1 
  and 2/3, respectively.}
  \label{rh-pericentre}
\end{minipage}
\end{figure*}

The full list of simulations performed is shown in Table \ref{results_table}. 
The UCD formation time is defined as the time when the change in half-mass radius and mass within 
the tidal radius \citep[calculated according to][]{King:1962} of the resulting object falls below 
$10\%$ between successive passages, while the final half-light radius and mass of the object is 
calculated at the end of the simulation (column 10 in Table \ref{results_table}). 
Since the change in the model is small at the end of the simulation (see Fig. \ref{sb_profile}) we 
can be confident that a longer simulation time will not significantly change our results. 
For the elliptic orbits the number of pericentre passages is the number before formation occurs and 
not the total number during the simulation.

%%%%%%%%%%%%%%%%%%%%%%%%%%%%%%%%%%%%%%%%%%
%                RESULTS                 %
%%%%%%%%%%%%%%%%%%%%%%%%%%%%%%%%%%%%%%%%%%
\section{RESULTS}

%%%%%% Section 3.1
\subsection{Evolution of galaxy size}

Fig. \ref{rh-pericentre} shows a comparison of the final half-light radii for the simulated UCDs. 
Except for models on an elliptic orbit with a 2 kpc pericentre, all simulated UCDs have sizes larger 
than the initial model nucleus. 
All simulations, except for models on a box orbit with a 100 kpc apocentre and two pericentre 
passages at 5 kpc (simulations 26, 34 and 42), form objects with half-light radii less than 100 pc. 
For both elliptic and box orbits the half-light radius increases with pericentre distance since 
models with a small pericentre distance suffer more tidal stripping than those with a large 
pericentre distance. 
For a given pericentre distance, 100 kpc apocentres tend to produce larger half-light radii than 50 
kpc apocentres because particles require more energy to escape the dwarf galaxy at a large 
galactocentric radius than at a small radius. 
Models on box orbits produce both larger half-light radii and a larger range of half-light radii 
than those on elliptic orbits ($10 \lesssim r_h / \rmn{pc} \lesssim 170$ for box orbits compared to 
$4 \lesssim r_h / \rmn{pc} \lesssim 25$ for elliptic orbits) due to box orbits having fewer close 
pericentre passages and therefore suffering less tidal stripping. 
In the box orbit scenario, an increase in the number of close pericentre passages results in a 
decrease in half-light radius due to the inner region of the model becoming more susceptible to 
tidal effects once the outer region has been removed. 

\begin{figure}
  \centering
  \includegraphics[width=84mm]{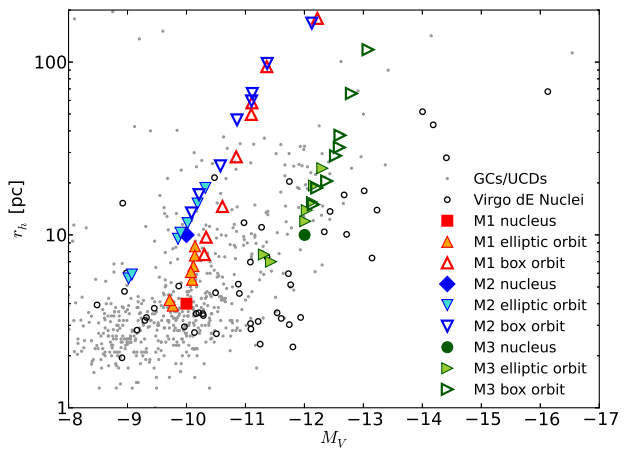}
  \caption{Final \textit{V}-band magnitude and half-light radius for the simulated UCDs compared 
  with GCs and UCDs from the nearby universe \citep[grey points]{Brodie:2011}, and Virgo dwarf 
  elliptical nuclei \citep[open black circles]{Cote:2006}. 
  The nuclei are converted to $V$-band photometry from $g$ and $z$ bands using the relation derived 
  for M87 globular clusters by \citet{Peng:2006}. 
  Symbols and colours are as in the legend, where M1, M2 and M3 are dE,N models 1, 2, and 3 runs, 
  respectively, and the original model nuclei sizes are represented by a red square (M1), blue 
  diamond (M2), and green circle (M3).
  The simulations are converted to a luminosity assuming a mass-to-light ratio $\Upsilon = 3$ 
  $(\rmn{M}/\rmn{L}_V)_\odot$.}
  \label{rh-MV}
\end{figure}

\begin{figure*}
\centering
\begin{minipage}{168mm}
  \centering
  \includegraphics[width=83.5mm]{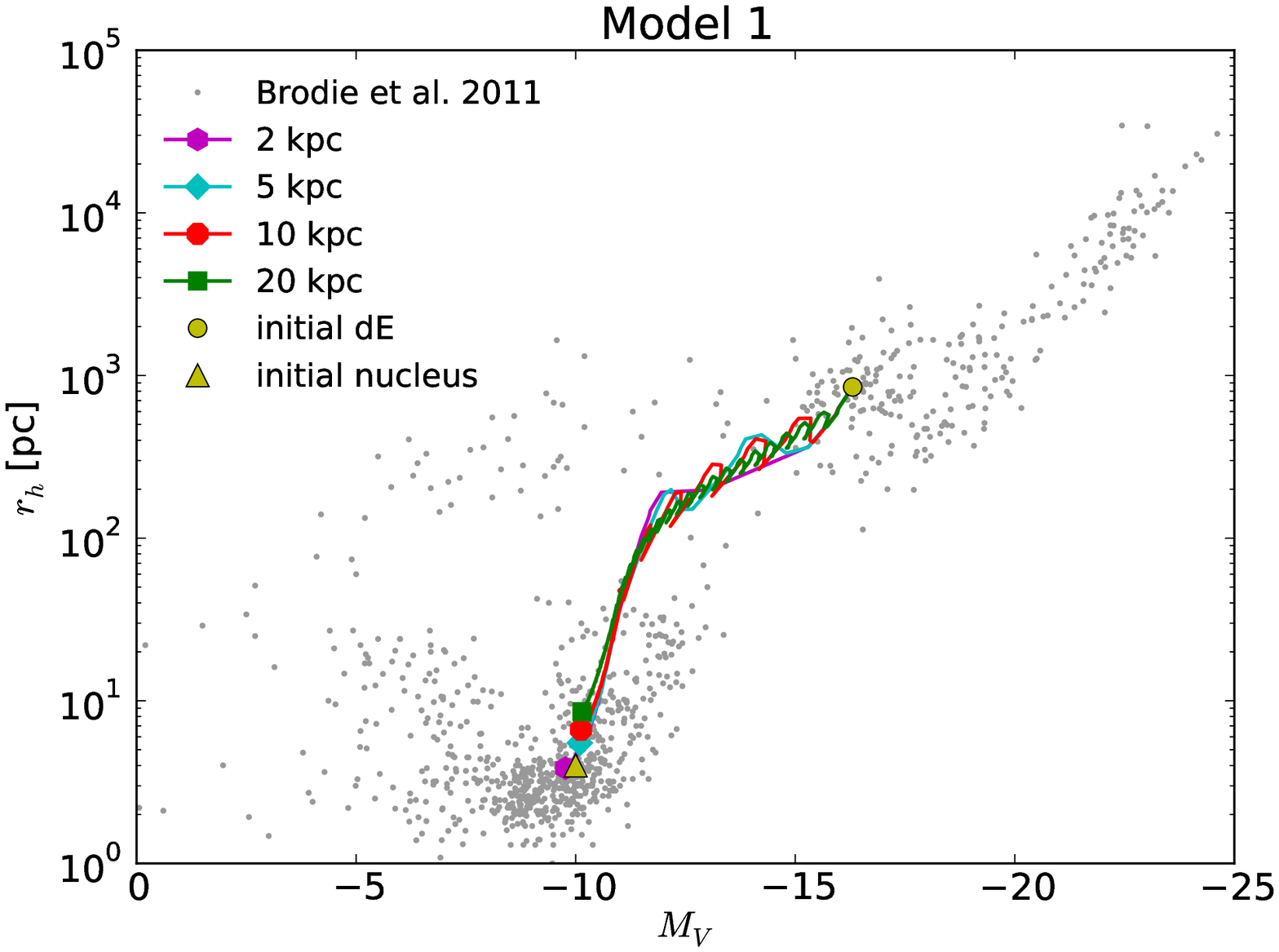}
  \includegraphics[width=83.5mm]{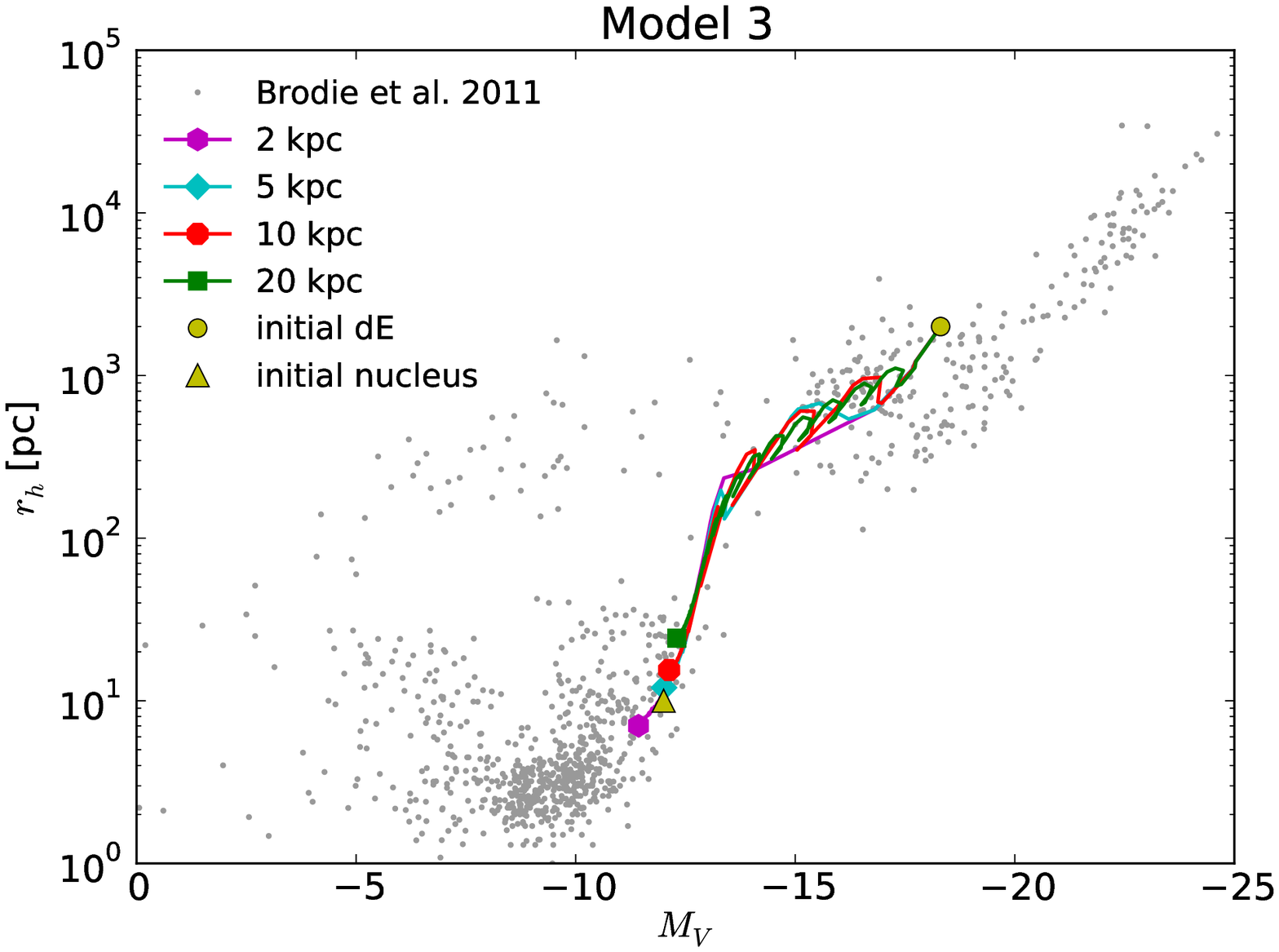}
  \caption{Morphology evolution of simulations on elliptic orbits with a 50 kpc apocentre for dE,N 
  model 1 (left panel) and dE,N model 3 (right panel) compared to GCs, UCDs, dEs, dwarf spheroidals,   compact ellipticals, and giant ellipticals from the nearby universe 
  \citep[grey points]{Brodie:2011}. 
  The \textit{V}-band magnitude and half-light radius evolution of the models are shown as solid 
  lines with a marker showing final UCD at the end of the simulation. 
  The marker type and line colour for each simulation are as in the legend.
  The yellow circle and triangle mark the initial position of the dwarf galaxy and nucleus of the 
  model, respectively.}
  \label{morphology}
\end{minipage}
\end{figure*}

Comparison of \textit{V}-band magnitude and half-light radius between the simulated UCDs and 
observed GCs, UCDs and dE nuclei is shown in Fig. \ref{rh-MV}. 
{Note that the final UCD in the simulations is a combination of the remaining particles from both 
the nucleus and the envelope and therefore can be more massive than the initial nucleus.}
For dE nuclei with $M_V > -11$ the median half-light radius is 3.5 pc, and thus our model with a 
nucleus $r_h = 4$ pc (model 1) is more comparable to present day nuclei than the model with 
$r_h = 10$ pc (model 2). 
For dE nuclei with $M_V < -11$ the median half-light radius is 7.5 pc and thus comparable to the 
model with a nucleus $M_V = -12$ and $r_h = 10$ pc (model 3). 
In the elliptic orbit scenario the models have a maximum final half-light radius approximately less 
than two times the original nucleus size, and therefore can only match observed UCDs if dE nuclei 
have initial half-light radii $r_h \gtrsim 10$ pc. 
In principle the extended UCDs could be formed on elliptic orbits with pericentres larger than 20 
kpc, however as formation time scales exponentially with pericentre distance, while UCD half-light 
radius scales as a power law with exponent $\sim 0.5$ with pericentre distance, the formation time 
required becomes much larger than a Hubble time for a dE,N with a nucleus size of $r_h \sim 4$ pc. 
In contrast, box orbits can produce half-light radii up to 40 times initial size of the model 
nucleus, and can produce the full range of observed UCD half-light radii with a nucleus half-light 
radius of either 4 or 10 pc. 
For model 3 the UCD sequence closely matches the most extended and massive UCDs ($-14 < M_V < -12.5$ 
and $60 < r_h/\rmn{pc} < 100$). 

The morphological evolution from dE to UCD for models 1 and 3 on elliptic orbits is shown in Fig. 
\ref{morphology}. 
Model 2 simulations take a similar path in $r_h-M_V$ space as model 1 due to the models having an 
identical enclosed mass profile for the dE envelope. 
All simulations for a given dE,N model take a similar path in $r_h-M_V$ space, first towards lower 
mass and then towards lower radius, with only the end point along the sequence differing depending 
upon the orbital parameters. 
The figure shows that some objects classified as dwarf galaxies (i.e. objects with $r_h > 100$ pc) 
could have undergone some tidal stripping or are still being tidally stripped.

Figures \ref{rh-MV} and \ref{morphology} also show that our simulations predict the existence of 
objects with half-light radii between 50 and 200 pc and luminosities smaller than $M_V \sim -12$, 
while the catalogue from \citet{Brodie:2011} lacks such objects. 
It is unclear whether this is a real effect or due to selection effects or selection bias, but note 
the catalogue compiled by \citet{Bruens:2012} contains unconfirmed candidates in this region.
If the extended, low mass UCDs form by tidal stripping, one would expect to see transition objects 
in the size gap. 
The absence or small number of objects in this region might imply different situations for UCD 
formation by tidal stripping: either the formation time is very short, or the tidal stripping 
occurred long ago, i.e. when the cluster formed. 
Our results show that UCDs on box orbits, which are necessary to form the extended UCDs, have 
typical formation times of 1-2 Gyr (see Table \ref{results_table}), in agreement with the first 
scenario.
However, since most UCDs are $\sim 10$-11 Gyr old \citep{Evstigneeva:2007a, Francis:2012}, while 
most dE nuclei are a few Gyr old \citep{Paudel:2011}, this implies that most UCDs formed before the 
young dE nuclei were formed, otherwise one could expect to find more young UCDs.
But note that dE nuclei with young ages might not be young at all, they might have formed long ago 
but had ongoing star formation or new star formation events at recent times.
Therefore either of these points, or a combination of the two, may explain the lack of transition 
objects between dwarf galaxies and UCDs.

\begin{figure}
  \centering
  \includegraphics[width=84mm]{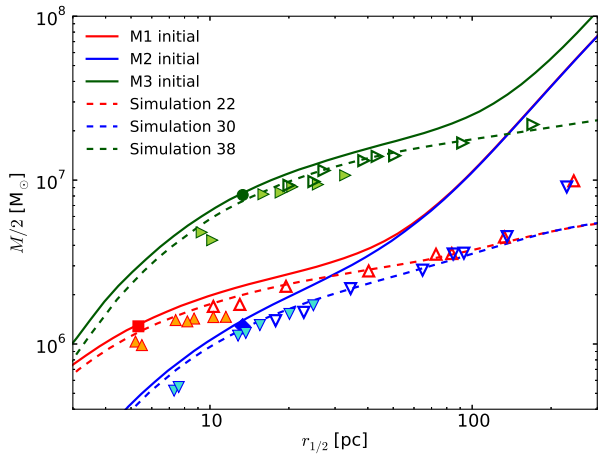}
  \caption{Final 3D half-mass radius $(r_{1/2})$ and half the object mass for all simulated UCDs 
  compared with the initial cumulative mass profiles for the models (solid lines) and final 
  cumulative mass profiles for simulations 22, 30 and 38 (box orbits with one pericentre passage at 
  2 kpc and a 50 kpc apocentre/circular orbit for models 1, 2 and 3 respectively, dashed lines). 
  Colours and symbols are as in Figure \ref{rh-MV}.}
  \label{half-mass}
\end{figure}

Fig. \ref{half-mass} shows a comparison of final half-mass radius and half the object mass for the 
simulations with the initial cumulative mass profile of the models and the final cumulative mass 
profile for box orbits with one pericentre passage at 2 kpc and a 50 kpc apocentre/circular orbit. 
If the origin of the size difference between UCDs and dE nuclei is due to expansion of the nucleus 
during tidal stripping the cumulative mass profiles of the final UCDs should differ from that of the 
initial nucleus of the model. 
Fig. \ref{half-mass} shows that the final UCD profiles differ little from nucleus of the initial 
models, and the mass and half-mass radius of the simulated UCDs trace the initial model nucleus, 
with the exception of the most massive UCDs for each model which have a significant stellar halo. 
This indicates expansion plays little role in the final UCD sizes. 
Although the model 1 elliptic orbit simulations show some deviation from the mass profile with 
increasing radius, this is most likely caused by numerical relaxation and not a real effect. 
Based on this result we expect that UCDs outside our sequence can be formed by models with different 
luminosities for the nucleus. 
Given the range of luminosities for dE nuclei in Fig. \ref{rh-MV}, the whole range of UCD sizes and 
luminosities could be produced by dE,N on box orbits.

%%%%%% Section 3.2
\subsection{UCD surface brightness profiles}

\begin{figure*}
\centering
\begin{minipage}{168mm}
  \centering
  \includegraphics[width=83.5mm]{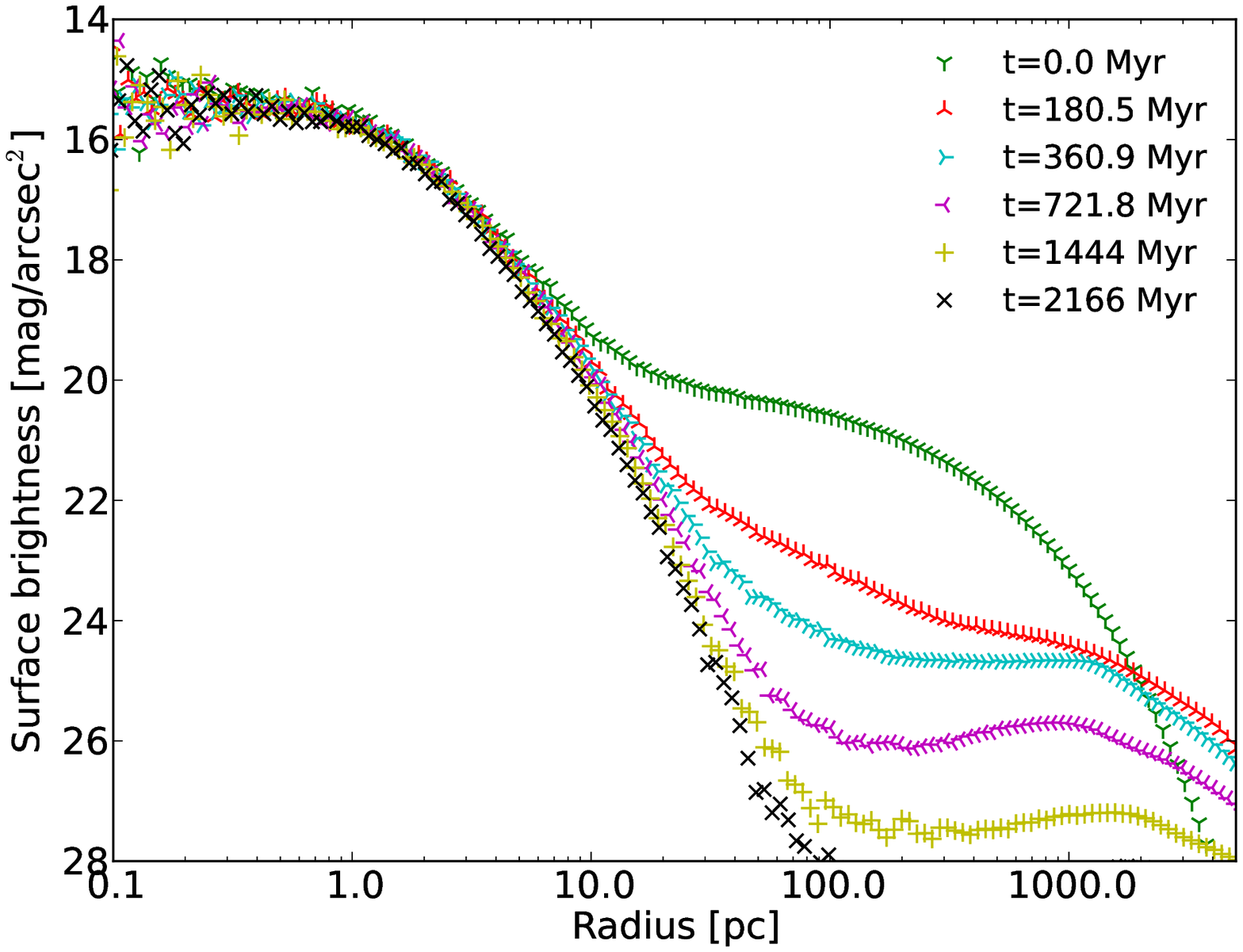}
  \includegraphics[width=83.5mm]{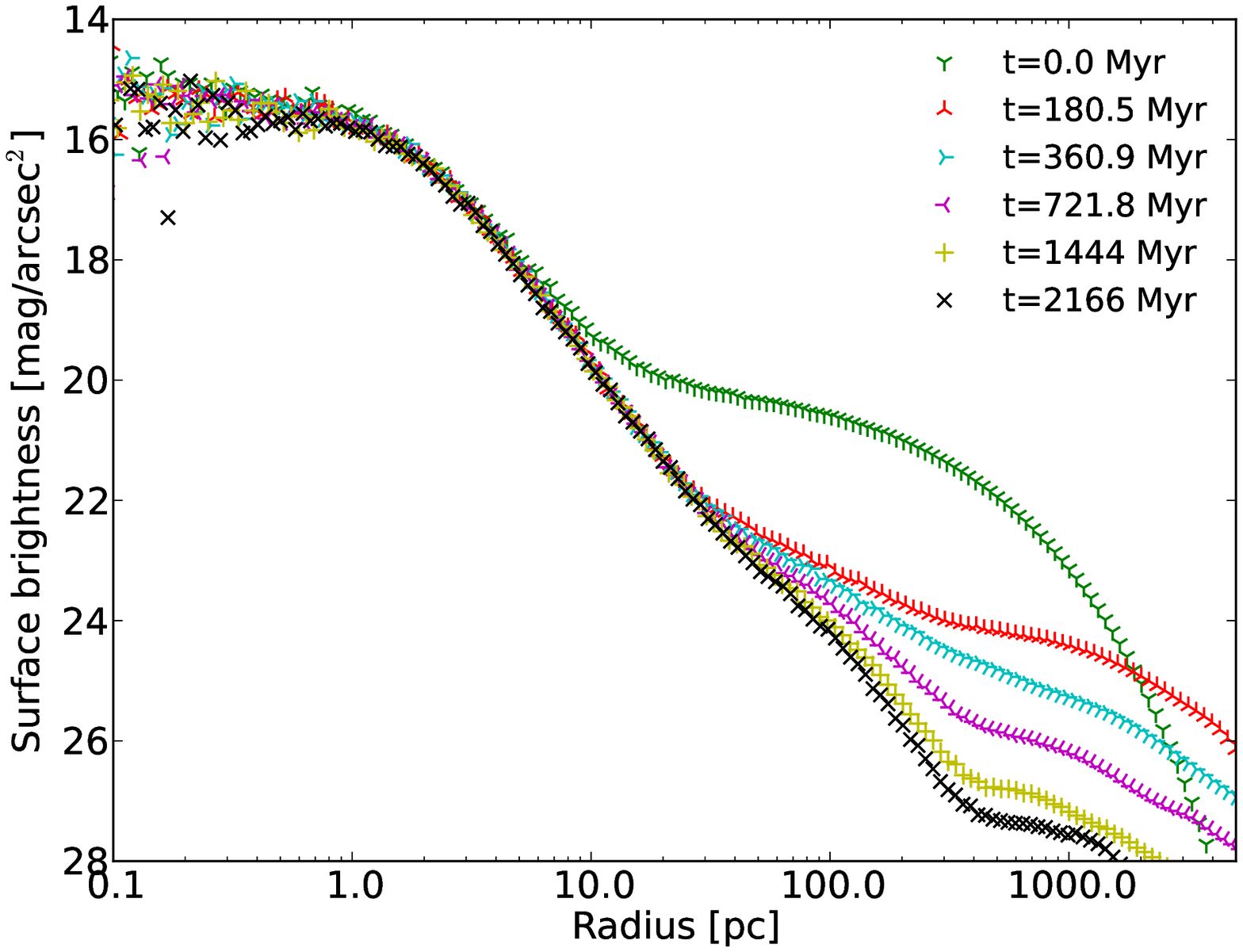} \\
  \caption{Surface brightness profile evolution over time for dE,N model 1. 
  \textit{Left.} An elliptic with a 2 kpc pericentre and 50 kpc apocentre (Simulation 1 in Table 1, 
  final $r_h = 3.9$ pc, $M_V = -9.77$). 
  \textit{Right.} A box orbit with one pericentre passage at 2 kpc and a 50 kpc apocentre/circular 
  orbit (Simulation 22 in Table 1, final $r_h = 28.3$ pc, $M_V = -10.84$). 
  The surface brightness profile symbols and colours are as in the legend, showing the time since 
  the start of simulation. }
  \label{sb_profile}
\end{minipage}
\end{figure*}

As an example, Fig. \ref{sb_profile} shows the surface brightness evolution for dE,N model 1 on an 
elliptic orbit with a 2 kpc pericentre and a 50 kpc apocentre (left panel) compared to the same 
model on a box orbit which has one close pericentre passage at 2 kpc and then continues on a 
circular orbit at 50 kpc (right panel). 
For the elliptic orbit, the dE,N is strongly stripped by the first two pericentre passages at 90 and 
270 Myr and the final UCD has a size and mass resembling the initial nucleus of the model. 
For the box orbit, most of the envelope becomes unbound due to the pericentre passage at 90 Myr and 
subsequently escapes during the orbit at apocentre, while the nucleus remains unaffected by the 
encounter. 
The final mass of the object is approximately twice that of the initial nucleus. 
Since our set-up procedure does not distinguish between nucleus and envelope particles we are unable 
to determine which particles escape, however, naively one would expect the envelope particles to 
become unbound before the nucleus particles. 
Under this assumption all particles in the nucleus would be retained while 50 per cent of the final 
UCD mass comes from the envelope.

In the right panel of Fig. \ref{sb_profile} the final profile has a stellar halo, which is visible 
as a deviation from a single component profile at a surface brightness of $\sim 23$ mag 
arcsec$^{-2}$. 
In general, for models with a nucleus $r_h = 4$ pc (model 1) stellar haloes become visible for UCDs 
with sizes larger than $r_h \sim 30$ pc, while for models with a nucleus $r_h = 10$ pc (model 2 and 
3) stellar haloes become visible for UCDs with sizes larger than $r_h \sim 50$-60 pc. 
For all simulations the halo does not become apparent until a surface brightness of 22-23 mag 
arcsec$^{-2}$. 
For simulation 22 (right panel in Fig. \ref{sb_profile}), it is unclear whether a halo would
be observable because if random Gaussian noise with a standard deviation of $0.2$ mag arcsec$^{-2}$ 
is added to simulate observational uncertainties, the profile is fit well by both a King profile and 
a two-component profile (King profile with a S{\'e}rsic profile for the halo). 
This indicates that some UCDs which only have a single component surface brightness profile could be 
composed of the nucleus and remaining envelope from the initial dE,N.
However, for the more massive and extended simulated UCDs the stellar halo is clearly visible.

The result that UCDs formed by tidal stripping are composed of stars from both the nucleus and 
envelope of the progenitor dE,N  has an important implication. 
Nuclei of dE,Ns often have different metallicities from the envelope \citep{Paudel:2011}, which 
suggests that UCDs with two-components, as well as some that have only a single component profile 
(e.g. right panel in Fig. \ref{sb_profile}), most likely contain populations with different 
metallicities or a metallicity gradient.
This prediction may be tested with future observations.

Tidal streams are an inevitable consequence of the tidal stripping scenario and therefore must be 
present if UCDs form in this way. 
In the right panel of Fig. \ref{sb_profile} the final profile is embedded in a tidal stream, which 
becomes dominant in the surface brightness profile at a surface brightness of $\sim 27$ mag 
arcsec$^{-2}$ and a radius of $\sim 400$ pc. 
For all simulations the tidal streams have surface brightnesses fainter than $\sim 27$ mag 
arcsec$^{-2}$ after 2 Gyr. 
Only box orbit simulations with 50 kpc apocentres have tidal streams with surface brightnesses 
brighter than 28 mag arcsec$^{-2}$ in the final profiles, while all other simulations have tidal 
streams with surface brightnesses fainter than 28 mag arcsec$^{-2}$. 
It is probable the tidal streams would disperse faster in reality due to substructure in the galaxy 
cluster and thus tidal streams are only likely to be observed around UCDs still undergoing 
significant stripping.

For both the elliptic and box orbit in Fig. \ref{sb_profile} the inner region (within $\sim 5$-10 
pc) changes little throughout the simulation. 
By fitting a King profile to the innermost region of the final surface brightness profiles for all 
models we find the core radius typically changes by less than 10 per cent of the initial core radius 
of the nucleus (dE,N model 1 has a nucleus core radius of 1.5 pc while dE,N models 2 and 3 both have 
a nucleus core radius of 3.7 pc). 
This suggests the surface brightness profiles of UCDs formed by tidal stripping should have the same 
core radius as their progenitor dE nucleus.

%%%%%%%%%%%%%%%%%%%%%%%%%%%%%%%%%%%%%%%%%%
%               DISCUSSION               %
%%%%%%%%%%%%%%%%%%%%%%%%%%%%%%%%%%%%%%%%%%
\section{DISCUSSION}

We have performed simulations of nucleated dwarf galaxies undergoing tidal stripping in a Virgo-like 
galaxy cluster to form UCDs. 
Using the particle-mesh code \textsc{superbox}, we have performed 45 simulations with varying 
orbital parameters to test the effect of the dwarf galaxy's orbit on the size of the UCD formed due 
to the stripping of the dwarf's outer envelope. 

We find that repeated close passages which occur during elliptic orbits lead to the formation of 
compact star clusters/UCDs, and elliptic orbits can only reproduce the full range of UCD sizes if 
the dE nuclei have half-light radii $r_h > 10$ pc. 
Given a large fraction of dE nuclei have half-light radii $r_h \sim 4$ pc (including almost all 
nuclei with luminosities $M_V > -11$), we consider tidal stripping on elliptic orbits unlikely to be 
the dominant mechanism for UCD formation. 
In contrast orbital motion in box orbits, or other orbits where very close pericentre passages 
happen only once or twice and at all other times the stripped dwarf galaxy is far from the centre of 
a major galaxy, lead to the formation of extended objects resembling UCDs regardless of the nucleus 
half-light radius. 
For such box orbits the dwarf galaxies need close pericentre passages with distances less than 10 
kpc to undergo strong enough stripping so that UCD formation is possible.

Observations suggest that the nuclei must expand by a factor of two to account for the size 
difference between UCDs and dE nuclei \citep{Evstigneeva:2008}, however we find the nuclei expand 
little during the tidal stripping process. 
Instead, the stripped dE,N galaxies can resemble the extended UCDs by retaining more mass than 
contained within the initial nucleus, causing the UCDs to be more extended.
During the stripping process the envelope profile steepens, and in some cases the UCD can appear to 
have a single component profile despite being composed of both the nucleus and remnant envelope.
For all orbits considered in our simulations, the typical UCD size is 2-3 times the initial dE 
nucleus size, in agreement with the findings of \citet{Evstigneeva:2008}.

Despite the simulated UCDs having more mass than the initial nucleus, we find only the extended 
UCDs have stellar haloes. 
For the dE,N models with a compact nucleus ($r_h = 4$ pc) we find haloes only become visible for 
objects with a half-light radius greater than $\sim 30$ pc, while for dE,N models with a larger 
nucleus ($r_h = 10$ pc) a halo becomes visible for a half-light radius larger than $\sim 50$-60 pc. 
For both nuclei sizes the halo tends to become visible in the surface brightness profile at 
$\sim 22$-23 mag arcsec$^{-2}$ and between $\sim 10$-100 pc, however, the position this occurs at 
for a given UCD will depend strongly on the mass profile of the progenitor dE,N (in particular on 
the King concentration for the nucleus and Sersic index for the envelope) and remaining mass in the 
envelope.
Some observed UCDs, in particular the most extended ones, are better fit by two-component models 
\citep{Evstigneeva:2008}, while deviations from single-component King models at a surface 
brightness of $\sim 22$-23 mag arcsec$^{-2}$ can be seen qualitatively in some of the observed 
surface brightness profiles of UCDs (e.g. the profiles for UCD16 and UCD33), in agreement with our 
results.

Given our results suggesting the nuclei undergo little expansion and retain the core radius of their 
progenitor dE nucleus, we predict that high mass UCDs ($M_V < -11$) should have cores ranging up to 
$\sim 20$ pc due to the large range in nuclei sizes at higher luminosities, assuming most progenitor 
nuclei have half-light radii below 20 pc. 
In contrast, most low mass UCDs should have profiles with cores up to a few parsec, since most 
nuclei in this range have half-light radii of $\sim 4$ pc. 
This range is consistent for the UCDs in the Fornax and Virgo clusters which have King core 
radii in the range 2-7 pc \citep{Evstigneeva:2008}. 
No surface brightness profiles are available for the extended, low luminosity UCDs ($r_h \sim 
40$ pc, $M_V \sim -9$) as yet but this prediction may be tested with future observations.

Although we simulated a dE with a nucleus of $M_V=-12$ and $r_h=10$ pc (model 3) there also exist 
many compact nuclei with $M_V=-12$ and $r_h=4$ pc, as shown in Fig. \ref{rh-MV}.  
If the envelope for both dE,Ns is similar, our results from Fig. \ref{half-mass} imply a dE,N with a 
nucleus of $M_V=-12$ and $r_h=4$ pc will evolve during tidal stripping in a similar way to model 3, 
up until $r_h \sim 100$ pc when the nucleus starts to become dominant in the mass profile. 
After this point the UCD would be more compact than model 3 for a given luminosity, similar to the 
situation between models 1 and 2.
The most massive nuclei in Fig. \ref{rh-MV} have $M_V<-14$ and $r_h \sim 40$ pc, however no UCDs are 
observed at these sizes. 
Our simulations show that tidally stripped dE,Ns resulting in UCDs less massive and less extended 
than the initial nucleus requires many (more than 5) pericentre passages with distances less than 5 
kpc.
We consider such orbits with many close pericentre passages unlikely since a small perturbation 
could increase the pericentre distance.
As the typical UCD size is 2-3 times the initial nucleus size, UCDs formed from such massive nuclei 
would likely have sizes of $r_h \sim 100$ pc and luminosities of $M_V \sim -15$.
Since the host galaxies of the most massive nuclei are lenticular (S0) or elliptical galaxies, 
and therefore more massive than dEs, the number of close pericentre passages required
to tidally strip the galaxy is likely much larger.
The extreme orbits required for formation, and the rarity of objects with such massive nuclei, may
explain the lack of objects in this region.
An extremely massive nucleus with $M_V=-16$ would most likely resemble a compact elliptical galaxy, 
rather than a UCD, after tidal stripping.

Two questions that remain unanswered are whether orbits with only one to two close passages required 
for extended UCD formation occur in real galaxy clusters, and whether dwarf galaxies are destroyed 
in sufficient numbers to explain all UCDs. 
For the first question a possible scenario in which such orbits may occur is if the first few 
passages of a dwarf galaxy in a galaxy cluster are highly radial, due to infall on low-angular 
momentum orbits, after which the pericentre increases due to a triaxial potential or interactions 
with other galaxies. 
A detailed answer to the first question, however, requires following the orbits of dwarf galaxies in 
cosmological simulations and we defer this, along with more accurate predictions of UCD numbers and 
spatial distributions, to a future paper. 

For the second question, there already exists some previous work constraining the number of possible 
progenitor galaxies for UCDs.
\citet{Mieske:2012} compared the specific frequencies of GCs and UCDs around various clusters and 
found no more than 50 per cent of UCDs may be formed by tidal stripping (based on the error bars of 
the specific frequencies derived for GCs). 
In the central $\sim 50$-70 kpc of the galaxy clusters they find $\gtrsim 90$ per cent of possible progenitor 
dwarfs need to be disrupted to account for half of the UCDs.
As tidal stripping will be most efficient near the centre of a galaxy cluster we consider it 
entirely possible that such a high fraction of dwarfs in this region have been tidally disrupted.

To date there have been 34 UCDs discovered in M87, with possibly more than 50 still undiscovered
out to a distance of 200 kpc \citep{Brodie:2011}. 
Following \citet{Mieske:2012}, we assume 50 per cent of the UCDs, $\sim 40$, to be formed by 
tidal stripping.
According to the catalogue of \citet*{Binggeli:1985}, about 50 dEs are located within a 
projected distance of 200 kpc from the centre of M87 \citep{Peng:2008}. 
Given approximately 70 per cent of dEs are nucleated \citep{Cote:2006, Turner:2012}, this 
leaves 35 dE,Ns to be possible UCD progenitors.
Therefore, approximately $\frac{40}{40+35} \approx 53$ per cent of possible UCD progenitor galaxies 
within 200 kpc of M87 need to be tidally stripped to account for half of the UCDs.
It is possible that observations are consistent with a population in which all dwarf galaxies are 
nucleated \citep*{Thomas:2008}, which would lower the fraction needed to be stripped by 10 per cent.
This estimate is consistent with calculations from cosmological simulations which suggest half of 
satellite galaxies get disrupted and/or accreted to their host halos \citep*{Henriques:2008}. 

Alternatively, one could obtain an estimate of accreted dwarf galaxy numbers by looking at the 
globular cluster systems of elliptical galaxies.
Giant elliptical galaxies contain very rich GC systems which are almost universally bimodal in the 
colour distributions due to differences in metallicity \citep{Brodie:2006}. 
Dwarf elliptical galaxies on the other hand contain GC systems which are predominantly metal-poor 
\citep{Peng:2008}. 
One explanation for this bimodallity of giant elliptical GCs is that the metal-rich GCs are the 
intrinsic GC population of the galaxy, or were formed in starbursts triggered by gas-rich mergers, 
while the metal-poor GCs are provided by accretion of dwarf galaxies 
\citep[e.g. see the review by][]{Richtler:2012}. 
This explanation is strengthened by current theories of giant elliptical formation where the 
dominant growth mechanism for the galaxies from $z=1$ to $z=0$ is accretion through minor mergers 
\citep*{Naab:2009}.
Two previous studies have investigated the build-up of the GC systems of giant ellipticals via 
accretion of galaxies: \citet*{Cote:1998} for NGC 4472, and \citet*{Hilker:1999b} for NGC 1399, 
although using slightly different methods.
These techniques provide a useful way to study the accretion of dwarf galaxies by giant ellipticals,
however such an analysis is beyond the scope of this paper.

A common origin of accretion and tidal stripping of dwarf galaxies for UCDs and blue GCs in giant 
ellipticals also places a constraint on the spatial distribution of UCDs. 
If this scenario is correct, due to their common origin, we expect UCDs far from the centre of M87 
to have a similar spatial distribution as the blue GCs. 
However at small distances UCDs should be underrepresented compared to GCs due to the ongoing tidal 
stripping converting UCDs into GC-like objects.
Some evidence for a more extended distribution of UCDs compared to GCs in the inner regions of 
galaxy clusters has been found \citep{Hilker:2011, Mieske:2012}, in agreement with our 
expectations.

In summary, we have demonstrated that the extended, low mass UCDs found by \citet{Brodie:2011} can 
be formed in the tidal stripping scenario providing the dwarf galaxies have one or two pericentre 
passages with distances less than 10 kpc. 
Given the observed range of dE nuclei sizes and luminosities, the full range of UCD sizes can be 
produced within the tidal stripping scenario.

%%%%%%%%%%%%%%%%%%%%%%%%%%%%%%%%%%%%%%%%%%
%            ACKNOWLEDGEMENT             %
%%%%%%%%%%%%%%%%%%%%%%%%%%%%%%%%%%%%%%%%%%
\section*{ACKNOWLEDGEMENTS}

We wish to thank Manuel Metz for providing a GPU enabled version of \textsc{superbox}, and Michael 
Drinkwater and the anonymous referee for helpful comments which improved the paper. 
JP is supported by the University of Queensland via an Australian Postgraduate Award and UQ 
Advantage RHD Scholarship. 
HB is supported by the Australian Research Council through Future Fellowship grant FT0991052 and 
Discovery Project grant DP110102608. 
Part of this work was performed on the gSTAR national facility at Swinburne University of Technology.
gSTAR is funded by Swinburne and the Australian Government’s Education Investment Fund.

%%%%%%%%%%%%%%%%%%%%%%%%%%%%%%%%%%%%%%%%%%
%              Bibliography              %
%%%%%%%%%%%%%%%%%%%%%%%%%%%%%%%%%%%%%%%%%%

\label{lastpage}

\end{document}